\documentclass[11pt,onecolumn]{ieeeconf}  
\usepackage{graphicx}
\usepackage[hidelinks]{hyperref}
\usepackage{subcaption}
\usepackage{amsfonts,amssymb}
\usepackage{tikz}
\usepackage{xcolor}

\IEEEoverridecommandlockouts                      

\usepackage{amsmath} 
\title{\LARGE \bf
Understanding Collective Stability of ACC Systems:\\ From Theory to Real-World Observations}

\author{Raphael Korbmacher$^{1}$, Parthib Khound$^{2}$ and Antoine Tordeux$^{1}$
\thanks{*This work was not supported by the German Research Foundation (DFG, grant number 546728715)}
\thanks{$^{1}$R.\ Korbmacher and A.\ Tordeux are with the chair of Traffic Safety and Reliability, University of Wuppertal, Wuppertal, Germany. Email addresses: {\tt korbmacher@uni-wuppertal.de} and {\tt tordeux@uni-wuppertal.de}.
$^{2}$P. Khound is with the Chair of Reliability of Technical Systems and Electrical Measurement, University of Siegen, Germany and the Electrical Engineering Department, Indian Institute of Technology Bombay, Mumbai, India. Email address: {\tt khoundparthib@gmail.com}}}

\begin{document}

\maketitle
\thispagestyle{empty}
\pagestyle{empty}

\begin{abstract}

Autonomated vehicles (AVs) are expected to have a profound impact on society, with high expectations for their potential benefits. One key anticipated benefit is the reduction of traffic congestion and stop-and-go waves, which negatively affect fuel efficiency, travel time, and environmental sustainability.
This paper presents a comprehensive review and meta-analysis of the impact of AVs on the longitudinal collective stability of vehicle single file. 
We focus on adaptive cruise control (ACC) systems, a widely used precursor technology to fully autonomous driving. ACC controllers have been studied extensively in both theoretical and practical contexts, making it a valuable starting point for analysis. Our study systematically differentiates the findings from models and simulations, controlled experiments, and empirical observations to provide a structured overview of existing research. Although some results in the literature are contradictory, three key insights emerge from our analysis: \textit{i}) String stability is highly dependent on the chosen time gap and reaction/response time.
\textit{ii}) ACC systems currently implemented in commercial vehicles are not string stable, as manufacturers prioritise individual comfort and smooth driving, resulting in high system response times.
\textit{iii}) Although cooperative ACC (CACC) systems are theoretically the most effective solution to ensure string stability, their widespread implementation in the near future remains uncertain. Instead, improvements in autonomous control algorithms should be considered to enhance system performance.

\end{abstract}
\medskip

\section{INTRODUCTION}
Adaptive Cruise Control (ACC) systems are fundamental components of autonomous driving. Similarly to car-following models, ACC systems consist of longitudinal dynamics controls based on the speed of current and preceding vehicles and the relative distance from the predecessor. These variables can be measured using sensors, typically a speedometer and a radar. The challenge is to formulate a dynamic control that provides safe, efficient, and comfortable behaviour for any longitudinal situation. Safety is generally conceived in terms of stability and oscillations (stop-and-go waves) or even collisions in dynamics. The robustness of the control to perturbations such as delays or noise is one of the other safety aspects to be considered. One of the first concepts of ACC systems dates back to the work of Ackermann in the early 1980s~\cite{ackermann1980abstandsregelung}. Supported by research projects such as the European Eureka Prometheus project, the first commercial applications appeared in the late 1990s. Today, most of the vehicle models available on the market offer ACC systems. The question of the strategy to be followed arose at an early stage of the research. In the mid-1990s, pioneering studies showed that the constant spacing strategy, which consists of maintaining a constant distance to the preceding vehicle, cannot achieve basic stability characteristics and must be ruled out, despite interesting efficiency features for platooning~\cite{swaroop1994comparision,rajamani2011vehicle}. In contrast, the \emph{Constant Time Gap} (CTG) strategy, where the distance to the predecessor is proportional to the current speed, shows interesting stability properties~\cite{swaroop1994comparision,zhou2005range,rajamani2011vehicle,zhou2004string}. The time gap is the time before a collision if the vehicle in front stops suddenly and if the speed of the ego-vehicle is kept constant. The strategy allows for the conservation of a constant physical reaction time with the predecessor, regardless of the speed, which is an interesting safety
characteristic. Maintaining a constant time gap is the strategy that is empirically observed for human drivers with a time gap between 1 and 2 seconds~\cite{banks2003average}. It is also the strategy
recommended by safety standards such as ISO 15622, with a time gap ranging from 0.8 to 2.2 seconds \cite{iso2018intelligent} or again 1.8 seconds for the German ADAC guidelines. The constant time gap is the most commonly used strategy in ACC systems. More sophisticated techniques are the \emph{Variable Time Gap} (VTG) strategies~\cite{wang2004should} or the quadratic range~\cite{zhou2005range}.

\section{Linear Stability Analysis}

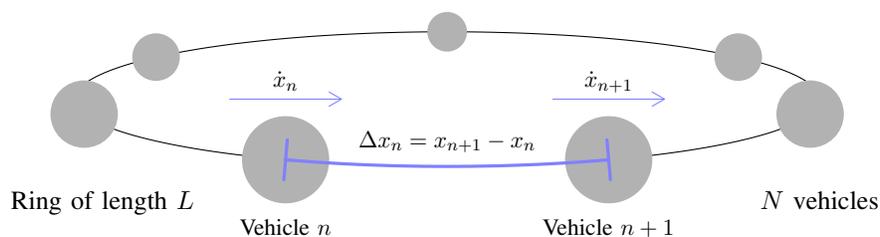
\begin{figure}[!ht]
    \centering




\begin{tikzpicture}[x=1.5pt,y=1.5pt]
\definecolor{fillColor}{RGB}{255,255,255}
\path[use as bounding box,fill=fillColor,fill opacity=0.00] (0,0) rectangle (247.08, 77.21);
\begin{scope}
\path[clip] ( 5.50, 5.50) rectangle (235.08, 65.21);
\definecolor{drawColor}{RGB}{0,0,0}

\path[draw=drawColor,line width= 0.4pt,line join=round,line cap=round] (123.54, 58.14) --
	(126.50, 58.14) --
	(129.46, 58.11) --
	(132.42, 58.07) --
	(135.36, 58.01) --
	(138.30, 57.93) --
	(141.22, 57.84) --
	(144.12, 57.73) --
	(147.00, 57.60) --
	(149.86, 57.46) --
	(152.69, 57.30) --
	(155.50, 57.13) --
	(158.27, 56.94) --
	(161.00, 56.73) --
	(163.70, 56.51) --
	(166.36, 56.27) --
	(168.98, 56.02) --
	(171.55, 55.75) --
	(174.07, 55.47) --
	(176.54, 55.18) --
	(178.96, 54.87) --
	(181.33, 54.54) --
	(183.64, 54.21) --
	(185.88, 53.86) --
	(188.07, 53.49) --
	(190.19, 53.12) --
	(192.24, 52.73) --
	(194.23, 52.34) --
	(196.14, 51.93) --
	(197.99, 51.51) --
	(199.76, 51.08) --
	(201.45, 50.63) --
	(203.06, 50.19) --
	(204.60, 49.73) --
	(206.05, 49.26) --
	(207.43, 48.78) --
	(208.72, 48.30) --
	(209.92, 47.81) --
	(211.04, 47.31) --
	(212.07, 46.81) --
	(213.01, 46.30) --
	(213.87, 45.79) --
	(214.63, 45.27) --
	(215.30, 44.75) --
	(215.88, 44.22) --
	(216.37, 43.69) --
	(216.77, 43.16) --
	(217.07, 42.63) --
	(217.28, 42.09) --
	(217.40, 41.56) --
	(217.42, 41.02) --
	(217.35, 40.48) --
	(217.19, 39.95) --
	(216.93, 39.41) --
	(216.58, 38.88) --
	(216.14, 38.35) --
	(215.60, 37.82) --
	(214.98, 37.30) --
	(214.26, 36.78) --
	(213.45, 36.26) --
	(212.55, 35.75) --
	(211.57, 35.25) --
	(210.49, 34.75) --
	(209.33, 34.25) --
	(208.08, 33.77) --
	(206.75, 33.29) --
	(205.34, 32.81) --
	(203.84, 32.35) --
	(202.27, 31.90) --
	(200.61, 31.45) --
	(198.88, 31.02) --
	(197.07, 30.59) --
	(195.20, 30.18) --
	(193.24, 29.77) --
	(191.22, 29.38) --
	(189.14, 29.00) --
	(186.98, 28.63) --
	(184.77, 28.27) --
	(182.49, 27.93) --
	(180.15, 27.60) --
	(177.76, 27.28) --
	(175.31, 26.98) --
	(172.82, 26.69) --
	(170.27, 26.42) --
	(167.67, 26.16) --
	(165.04, 25.91) --
	(162.36, 25.68) --
	(159.64, 25.47) --
	(156.88, 25.27) --
	(154.10, 25.09) --
	(151.28, 24.92) --
	(148.43, 24.77) --
	(145.56, 24.64) --
	(142.67, 24.52) --
	(139.76, 24.42) --
	(136.83, 24.34) --
	(133.89, 24.27) --
	(130.94, 24.22) --
	(127.98, 24.18) --
	(125.02, 24.17) --
	(122.06, 24.17) --
	(119.09, 24.18) --
	(116.14, 24.22) --
	(113.18, 24.27) --
	(110.24, 24.34) --
	(107.32, 24.42) --
	(104.40, 24.52) --
	(101.51, 24.64) --
	( 98.64, 24.77) --
	( 95.80, 24.92) --
	( 92.98, 25.09) --
	( 90.19, 25.27) --
	( 87.44, 25.47) --
	( 84.72, 25.68) --
	( 82.04, 25.91) --
	( 79.40, 26.16) --
	( 76.81, 26.42) --
	( 74.26, 26.69) --
	( 71.76, 26.98) --
	( 69.32, 27.28) --
	( 66.92, 27.60) --
	( 64.59, 27.93) --
	( 62.31, 28.27) --
	( 60.09, 28.63) --
	( 57.94, 29.00) --
	( 55.85, 29.38) --
	( 53.83, 29.77) --
	( 51.88, 30.18) --
	( 50.00, 30.59) --
	( 48.20, 31.02) --
	( 46.47, 31.45) --
	( 44.81, 31.90) --
	( 43.24, 32.35) --
	( 41.74, 32.81) --
	( 40.33, 33.29) --
	( 38.99, 33.77) --
	( 37.75, 34.25) --
	( 36.59, 34.75) --
	( 35.51, 35.25) --
	( 34.52, 35.75) --
	( 33.63, 36.26) --
	( 32.82, 36.78) --
	( 32.10, 37.30) --
	( 31.47, 37.82) --
	( 30.94, 38.35) --
	( 30.49, 38.88) --
	( 30.14, 39.41) --
	( 29.89, 39.95) --
	( 29.72, 40.48) --
	( 29.65, 41.02) --
	( 29.68, 41.56) --
	( 29.79, 42.09) --
	( 30.00, 42.63) --
	( 30.31, 43.16) --
	( 30.70, 43.69) --
	( 31.19, 44.22) --
	( 31.78, 44.75) --
	( 32.45, 45.27) --
	( 33.21, 45.79) --
	( 34.06, 46.30) --
	( 35.01, 46.81) --
	( 36.04, 47.31) --
	( 37.16, 47.81) --
	( 38.36, 48.30) --
	( 39.65, 48.78) --
	( 41.02, 49.26) --
	( 42.48, 49.73) --
	( 44.01, 50.19) --
	( 45.63, 50.63) --
	( 47.32, 51.08) --
	( 49.09, 51.51) --
	( 50.93, 51.93) --
	( 52.85, 52.34) --
	( 54.83, 52.73) --
	( 56.89, 53.12) --
	( 59.01, 53.49) --
	( 61.19, 53.86) --
	( 63.44, 54.21) --
	( 65.75, 54.54) --
	( 68.11, 54.87) --
	( 70.53, 55.18) --
	( 73.01, 55.47) --
	( 75.53, 55.75) --
	( 78.10, 56.02) --
	( 80.72, 56.27) --
	( 83.38, 56.51) --
	( 86.07, 56.73) --
	( 88.81, 56.94) --
	( 91.58, 57.13) --
	( 94.38, 57.30) --
	( 97.22, 57.46) --
	(100.07, 57.60) --
	(102.96, 57.73) --
	(105.86, 57.84) --
	(108.78, 57.93) --
	(111.71, 58.01) --
	(114.66, 58.07) --
	(117.61, 58.11) --
	(120.57, 58.14) --
	(123.54, 58.14);
\definecolor{fillColor}{gray}{0.70}

\path[fill=fillColor] ( 82.80, 25.85) circle ( 11);

\path[fill=fillColor] (164.27, 25.85) circle ( 11);

\path[fill=fillColor] (215.07, 37.37) circle ( 8.5);

\path[fill=fillColor] (196.94, 51.75) circle (  6);

\path[fill=fillColor] (123.54, 58.14) circle (  5);

\path[fill=fillColor] ( 50.13, 51.75) circle (  6);

\path[fill=fillColor] ( 32.00, 37.37) circle ( 8.5);
\definecolor{drawColor}{RGB}{128,128,255}

\path[draw=drawColor,line width= 1.2pt,line join=round,line cap=round] ( 82.80, 25.85) --
	( 83.57, 25.78) --
	( 84.34, 25.72) --
	( 85.12, 25.65) --
	( 85.90, 25.59) --
	( 86.68, 25.53) --
	( 87.46, 25.47) --
	( 88.25, 25.41) --
	( 89.04, 25.35) --
	( 89.83, 25.30) --
	( 90.63, 25.24) --
	( 91.43, 25.19) --
	( 92.23, 25.14) --
	( 93.03, 25.09) --
	( 93.84, 25.04) --
	( 94.65, 24.99) --
	( 95.46, 24.94) --
	( 96.27, 24.90) --
	( 97.09, 24.85) --
	( 97.91, 24.81) --
	( 98.73, 24.77) --
	( 99.55, 24.73) --
	(100.37, 24.69) --
	(101.20, 24.65) --
	(102.02, 24.62) --
	(102.85, 24.58) --
	(103.69, 24.55) --
	(104.52, 24.52) --
	(105.35, 24.49) --
	(106.19, 24.46) --
	(107.03, 24.43) --
	(107.86, 24.40) --
	(108.70, 24.38) --
	(109.55, 24.35) --
	(110.39, 24.33) --
	(111.23, 24.31) --
	(112.08, 24.29) --
	(112.92, 24.27) --
	(113.77, 24.26) --
	(114.61, 24.24) --
	(115.46, 24.23) --
	(116.31, 24.22) --
	(117.16, 24.20) --
	(118.01, 24.19) --
	(118.86, 24.19) --
	(119.71, 24.18) --
	(120.56, 24.17) --
	(121.41, 24.17) --
	(122.26, 24.17) --
	(123.11, 24.16) --
	(123.96, 24.16) --
	(124.82, 24.17) --
	(125.67, 24.17) --
	(126.52, 24.17) --
	(127.37, 24.18) --
	(128.22, 24.19) --
	(129.07, 24.19) --
	(129.92, 24.20) --
	(130.77, 24.22) --
	(131.62, 24.23) --
	(132.46, 24.24) --
	(133.31, 24.26) --
	(134.16, 24.27) --
	(135.00, 24.29) --
	(135.85, 24.31) --
	(136.69, 24.33) --
	(137.53, 24.35) --
	(138.37, 24.38) --
	(139.21, 24.40) --
	(140.05, 24.43) --
	(140.89, 24.46) --
	(141.72, 24.49) --
	(142.56, 24.52) --
	(143.39, 24.55) --
	(144.22, 24.58) --
	(145.05, 24.62) --
	(145.88, 24.65) --
	(146.71, 24.69) --
	(147.53, 24.73) --
	(148.35, 24.77) --
	(149.17, 24.81) --
	(149.99, 24.85) --
	(150.81, 24.90) --
	(151.62, 24.94) --
	(152.43, 24.99) --
	(153.24, 25.04) --
	(154.04, 25.09) --
	(154.85, 25.14) --
	(155.65, 25.19) --
	(156.45, 25.24) --
	(157.24, 25.30) --
	(158.04, 25.35) --
	(158.83, 25.41) --
	(159.62, 25.47) --
	(160.40, 25.53) --
	(161.18, 25.59) --
	(161.96, 25.65) --
	(162.73, 25.72) --
	(163.51, 25.78) --
	(164.27, 25.85);

\path[draw=drawColor,line width= 1.2pt,line join=round,line cap=round] ( 82.80, 25.85) -- ( 83.57, 25.78);

\path[draw=drawColor,line width= 1.2pt,line join=round,line cap=round] ( 83.24, 30.89) --
	( 82.80, 25.85) --
	( 82.37, 20.81);

\path[draw=drawColor,line width= 1.2pt,line join=round,line cap=round] (164.27, 25.85) -- (163.51, 25.78);

\path[draw=drawColor,line width= 1.2pt,line join=round,line cap=round] (164.71, 20.81) --
	(164.27, 25.85) --
	(163.84, 30.89);
\definecolor{drawColor}{RGB}{0,0,0}


\definecolor{drawColor}{RGB}{128,128,255}

\path[draw=drawColor,line width= 0.4pt,line join=round,line cap=round] ( 68.72, 41.15) -- ( 96.89, 41.15);

\path[draw=drawColor,line width= 0.4pt,line join=round,line cap=round] ( 93.76, 39.35) --
	( 96.89, 41.15) --
	( 93.76, 42.96);
\definecolor{drawColor}{RGB}{0,0,0}

\node[text=drawColor,anchor=base,inner sep=0pt, outer sep=0pt, scale=  0.8] at ( 82.80, 44.25) {{$\dot x_n$}};

\definecolor{drawColor}{RGB}{128,128,255}

\path[draw=drawColor,line width= 0.4pt,line join=round,line cap=round] (150.19, 41.15) -- (178.36, 41.15);

\path[draw=drawColor,line width= 0.4pt,line join=round,line cap=round] (175.23, 39.35) --
	(178.36, 41.15) --
	(175.23, 42.96);
\definecolor{drawColor}{RGB}{0,0,0}

\node[text=drawColor,anchor=base,inner sep=0pt, outer sep=0pt, scale=  0.8] at (164.27, 44.25) {{$\dot x_{n+1}$}};

\node[text=drawColor,anchor=base,inner sep=0pt, outer sep=0pt, scale=  0.8] at (123.54, 29) {{$\Delta x_n=x_{n+1}-x_n$}};

\node[text=drawColor,anchor=base,inner sep=0pt, outer sep=0pt, scale=  0.90] at ( 36.69, 13.45) {Ring of length {$L$}};

\node[text=drawColor,anchor=base,inner sep=0pt, outer sep=0pt, scale=  0.90] at (217.43, 13.45) {{$N$} vehicles};

\node[text=drawColor,anchor=base,inner sep=0pt, outer sep=0pt, scale=  0.80] at ( 82.80, 7) {Vehicle {$n$}};

\node[text=drawColor,anchor=base,inner sep=0pt, outer sep=0pt, scale=  0.80] at (164.27, 7) {Vehicle  {$n+1$}};

\end{scope}
\end{tikzpicture}

    \caption{Scheme of the vehicle systems on a one-dimensional circuit with periodic boundaries and notations of the main variables.}
    \label{fig:notations}
\end{figure}

The study of car-following behaviour, fundamental to understanding longitudinal vehicle dynamics, traces its origins to the 1950s with pioneering contributions from Reuschel~\cite{reuschel1950vehicle} and Pipes~\cite{pipes1953operational}. These early models captured vehicle interactions with their immediate predecessors using first- and second-order differential equations, initially neglecting time delays. Subsequent advancements incorporated delays into both linear and nonlinear frameworks, sparking the formal analysis of stability in car-following dynamics. 
The analysis is generally carried out on an infinite line or a one-dimensional circuit with periodic boundary as shown in Fig.~\ref{fig:notations}.

\begin{figure}[!ht]
    \centering
    \input{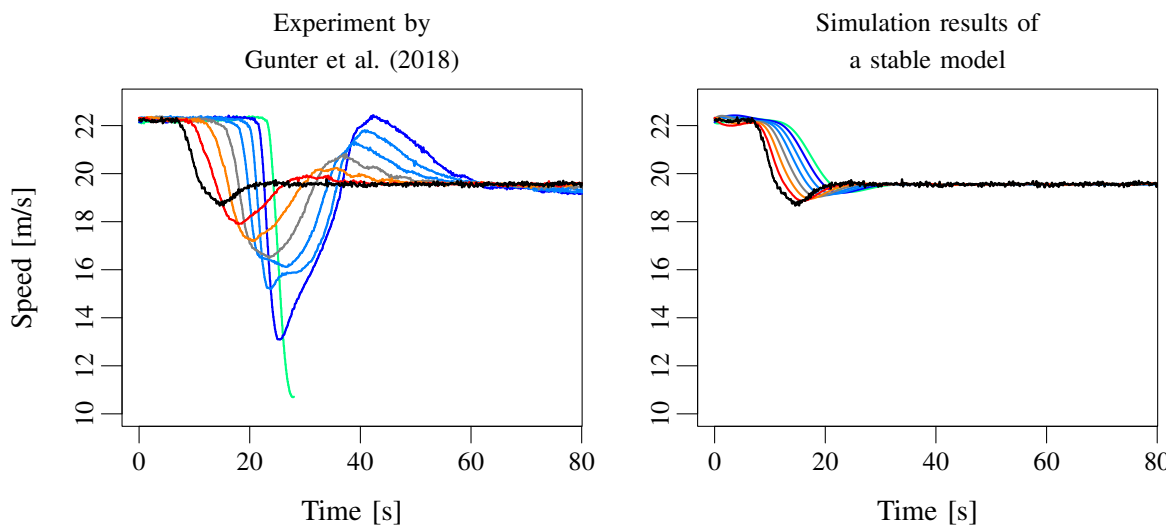}
    \caption{Velocity of an eight vehicle platoon. Left panel: showing an unstable platoon observed in the experiments from Gunter et al.~\cite{gunter2020commercially}. Right panel: for a simulated platoon using a stable CTG-ACC model. The trajectories can be computed and visualised online, see \cite{SimOverdamped}.}
    \label{fig:1} 
\end{figure}

In the context of traffic flow and vehicle control, stability is categorised into three key types \cite{treiber2013traffic}. The first one is \emph{local stability}. Having a leader $n+1$, the follower vehicle $n$ is local stable if the spacing error converges to zero when the leader travels at a constant speed $v$~\cite{khound2023extending}. This means that an ACC controller is locally stable, if it can adjust to small perturbations and return to its target speed and distance. 
Extending local stability to a finite group of vehicles, often led by a designated leader, so-called 
\emph{platoon stability}, where the collective response of a group to perturbations is examined. While mathematically akin to local stability across multiple vehicles, its practical implications differ due to inter-vehicle interactions. In Fig.~\ref{fig:1}, left panel, the speed-time profile of the ACC experiments of Gunter et al.~\cite{gunter2020commercially} are displayed. In contrast to the simulation results of the \emph{Constant Time Gap} (CTG) ACC model shown in Fig.~\ref{fig:1}, right panel, the ACC systems tested in the experiment demonstrated platoon instability, with perturbations amplifying upstream to the point where the last vehicle in the platoon has to significantly decrease the speed, causing the ACC system to disengage.

The last form of stability in the longitudinal behaviour of vehicles is \emph{string stability}. In this case, there is not a finite group of vehicles, but an infinite one and a finite system with periodic boundary conditions without a leader.
String stability conditions are stronger than local ones, as they take into account advective and convective perturbations that vanish locally.
A system is string stable if disturbances do not amplify as they stream through vehicle flow for general purpose~\cite{treiber2013traffic}. 
This is a critical consideration for the deployment of ACC systems on highways.

To formalize these concepts, consider the position of the $n$-th vehicle at time $t$ as $x_n(t)$, its velocity as $v_n(t)$, and the distance to the vehicle ahead as $\Delta x_n(t) = x_{n+1}(t) - x_n(t)$ (see Fig.~\ref{fig:notations}). A general second-order car-following model is expressed as:
\begin{equation}
\left\{~\begin{aligned}
    \dot x_n(t)&=v_n(t)\\[1mm]
    \dot{v}_n(t)&=F(\Delta x_n(t),v_n(t),v_{n+1}(t)),
\end{aligned}\right.
    \label{eq:ModGeneral}
\end{equation}
where $F$ dictates the acceleration based on spacing, the vehicle’s own speed, and the speed of the predecessor. 
For instance, the function for the Optimal Velocity Model (OVM) is given by
\begin{equation}
    F(\Delta x_n,v_n,v_{n+1})=\frac{1}{\tau}\big[ V\big(\Delta x_n\big) - v_n\big],
    \label{eq:OVM}
\end{equation}
where $\tau>0$ is the relaxation time and $V$ is the optimal velocity function. For the OVM, the speed of the predecessor does not influence the driver behaviour.
In any case, the car-following models have equilibrium solutions $(\Delta x^\star, v^\star) \in \mathbb{R}_+^2$ such that
\begin{equation}
    F(\Delta x^\star,v^\star,v^\star)=0.
\end{equation}
This equilibrium corresponds to a steady-state uniform condition in which all vehicles maintain a constant spacing $\Delta x^\star$ and velocity $v^\star$, resulting in zero acceleration. 
For instance, for the OVM with $N$ vehicles on a periodic circuit of length $L$, such steady states are given by $\Delta x^\star=L/N$ and $v^\star=V(L/N)$. Stability analysis examines how the system responds to small perturbations around this equilibrium, revealing whether it returns to the steady state (stable) or diverges (unstable).

Perturbations in spacing and velocity are defined as:
\begin{equation}
\begin{aligned}
    \Delta \tilde{x}_n(t) &= \Delta x_n(t) - \Delta x^\star, \\
    \tilde{v}_n(t) &= v_n(t) - v^\star,
\end{aligned}
\end{equation}
respectively.
The equilibrium solution $(\Delta x^\star, v^\star)$ is considered stable if these perturbations decay to zero as $t \to \infty$ for all $N$ vehicles~\cite{treiber2013traffic}. To analyse this, the model is linearised around the equilibrium by approximating $F$ locally using its partial derivatives and a first order Taylor expansion, yielding:
 \begin{equation}
     \dot{\tilde{v}}_n(t) = a\Delta\tilde{x}_n(t) + b\tilde{v}_n(t) + c\tilde{v}_{n+1}(t)\,,
 \end{equation}
where $a = \partial F / \partial \Delta x_n$, $b = \partial F / \partial v_n$, and $c = \partial F / \partial v_{n+1}$, evaluated at the equilibrium. These coefficients allow us to derive general stability conditions for any car-following model. In the following two subsections, we recall these conditions in the case of local stability and string stability.

\subsection{Local Stability}

Local stability ensures that perturbations in a single vehicle’s state decay when the leader maintains a constant speed. A stricter variant, overdamped local stability, mandates monotonic decay without oscillations, enhancing smoothness in ACC responses~\cite{khound2023extending}.
The local linear stability holds true as soon as $a>0$ and $b<0$ for any $c\in\mathbb R$. 
The overdamped local linear stability is more restrictive.
It requires~\cite{treiber2013traffic, tordeux2012linear}:
\begin{equation}
    a > 0, \qquad b<0,\qquad b^2\ge 4a.
    \label{eq:local_stability}
\end{equation}
For the OVM \eqref{eq:OVM}, assuming $V(x)=(x-\ell)/T$, where $T$ is the desired time gap parameter and $\ell\ge0$ is the length of the vehicle, we have $a = V'(\Delta x^\star)/\tau = 1/(T\tau)$, $b = -1/\tau$, and $c = 0$. Substituting this into \eqref{eq:local_stability}, the model is locally linearly stable if $\tau,T>0$ and overdamped stable if $0 < 4\tau \le T$. This last condition ensures a non-oscillatory decay of perturbations for a single vehicle. The local stability of different car-following models can be analysed online by simulation in \cite{SimLocal}.

\subsection{String Stability}
String stability requires that perturbations do not amplify globally, i.e., collectively, 
a more stringent condition than local stability. General string stability permits bounded oscillations, while overdamped string stability requires monotonic decay across the platoon, thereby guaranteeing the attenuation of transient effects. The explicit in depth study of the overdamped string stability criterion is relatively new, see e.g.,~\cite{khound2023unified}. This is a very strict stability condition that also guarantees collision avoidance during car-following manoeuvrers~\cite{lunze2018adaptive}.
For the linearised perturbation model, the general string linear stability conditions are  \cite{treiber2013traffic,tordeux2012linear}:
\begin{equation}
    a > 0, \qquad b < 0, \qquad b^2 - c^2 \ge 2a.
    \label{eq:string_stability}
\end{equation}
For the OVM where $a = 1/(T\tau)$, $b = -1/\tau$, and $c = 0$, these conditions simplify to $0 < 2\tau \le T$, i.e., the desired time gap must be two times larger than the relaxation time. This requirement is stricter than that for local linear stability, reflecting the challenge of maintaining collective stability, but less stringent than that for the local overdamped linear stability. Note that this last assertion does not systematically hold for models based on the speed of the predecessor. 
Achieving overdamped string stability requires additional constraints on the damping terms. For a platoon to achieve overdamped 
stability, particularly when~$c=0$, it is sufficient to satisfy the overdamped local stability criterion~\eqref{eq:local_stability}. 
For instance, the OVM \eqref{eq:OVM} is overdamped string stable if $0<4\tau \le T$. 
However, when~$c>0$, additionally the system should fulfil~\cite{khound2020local}: 
\begin{equation}
    \frac{a}{c}+b+c \le 0\,.
    \label{eq:string_stability_overdamped}
\end{equation}
Note that the overdamped string stability condition \eqref{eq:string_stability_overdamped} is stronger than the local overdamped condition \eqref{eq:local_stability} and the string condition \eqref{eq:string_stability}. 
In addition, the local overdamped condition \eqref{eq:local_stability} is stronger than the string condition \eqref{eq:string_stability} if $c^2\le 2a$. 
The string stability of different car-following models can be analysed online by simulation with a periodic system in \cite{SimString}.

\section{Meta-analysis}
The stability of ACC systems is extensively studied in the literature through simulations, as well as experiments under laboratory conditions and in the field. In addition, the benefits of cooperative systems that integrate vehicle communication are often compared with the performance of traditional autonomous ACC systems. 
In this section, we propose a meta-analysis to review the current state of research in this field.

\subsection{Stability of ACC Systems in Simulations}
The first market acceptance of ACC systems in passenger vehicles occurred in Japan in 1995, marking a significant milestone in automotive technology~\cite{bishop2005intelligent}. These systems quickly gained global popularity, prompting researchers to study their impact through simulations. In these simulation studies, assumptions about the behaviour of ACC systems were made, often relying on car-following models as the basis for analysis. The first simulation studies about the effect of ACC can be found in the DOMINC project~\cite{broqua1991cooperative}. Estimates suggest that traffic flow improves by 13\% when only 40\% of vehicles are equipped with ACC systems and the target time gap is set to 1 second. Marsden et al.~\cite{marsden2001towards} show significant improvements in comfort and environmental benefits. Treiber and Helbing~\cite{treiber2001microsimulations} use the \emph{Intelligent Driver} model (IDM)~\cite{treiber2000congested} to show how ACC systems can have tremendous positive effect on stability. They state that if only 20\% of vehicles are equipped with specific ACC controllers, they can eliminate congestion almost completely. Davis~\cite{davis2004effect} simulates mixed traffic by using modified OVMs for human drivers and the IDM for ACC. The results indicate that a 20\% ACC penetration rate could suppress traffic jams in the simulated environment. Bose and Ioannou~\cite{bose2001analysis} adopt the Pipes model~\cite{pipes1953operational} to simulate human drivers and a first-order model~\cite{ioannou1994throttle} for the ACC vehicles. Their simulations reveal smoother traffic flow and highlight positive environmental impacts, even with ACC penetrations as low as 10\%. Notably, they emphasise that ACC systems exhibit string stability in homogeneous conditions but struggle to maintain it in mixed traffic scenarios. 
Other relevant car-following models for ACC systems are extensions of the \emph{Intelligent Driver} and \emph{Full Velocity Difference} models \cite{bai2024novel,qin2023stabilizing}, the model of Newell \cite{zhou2022congestion,coppola2024context}, or again the \emph{Adaptive Time Gap} (ATG) model \cite{khound2023extending}. 
These models are all able to simulate string stable car-following scenarios for certain parameter settings. However, ACC systems in real situations have to handle challenges like time delays (also called reaction time), actuator lags, and also physical random factors such as road curvature, texture and slope, wind effects, lane-changing manoeuvres, etc. Incorporating these factors into models requires revisiting the question of stability from the ground up. For example, the predictor feedback model~\cite{bekiaris2020pde} utilises the finite spectrum assignment~(FSA) approach to address string stability issues caused by time delays, as demonstrated through simulation experiments. Similarly, the overdamped string-stable strategy compensating lag, delay, and disturbance (OSSCLDD)~\cite{khound2023unified}, theoretically addresses more of these challenges. Simulation results of the OSSCLDD model indicate overdamped string stability across various spacing strategies, such as CTG and VTG.

\subsection{Stability of ACC Systems in Experiments} 
\label{Experiements}

Analysing ACC systems in simulations is an important approach to better understand controller designs and parameter settings. Nevertheless, it is not sufficient to fully understand its true physical effects on traffic flow, especially in terms of string stability. Consequently, experiments with real ACC-equipped vehicles are essential. One reason is that the controllers of commercial ACC vehicles are not publicly available, and another reason is that experimental behaviour can be influenced by factors like sensing errors and delays \cite{li2021car}.
One of the first experiments on string stability was the circuit experiment conducted by Sugiyama et al., 2008~\cite{sugiyama2008traffic}. Their study demonstrates how a uniform flow of 22 vehicles on a closed-loop track could spontaneously transition into a traffic jam due to small fluctuations in driver behaviour, even in the absence of bottlenecks or external disruptions. This finding was reinforced by other circuit experiments, such as those conducted by Tadaki et al. 2013~\cite{tadaki2013phase}. Additional experiments have also confirmed the string instability of human drivers in real-world conditions~\cite{jiang2014traffic,jiang2018experimental}. It is also worth noting that the first practical demonstration of the ACC system in a platoon was conducted by the California Partners for Advanced Transit and Highways (PATH) program in August 1997, proving the technical feasibility of such a system \cite{rajamani2011vehicle, shladover2007path}.

To better understand the potential benefits of ACC systems, similar experiments have been conducted, incorporating both human-driven and ACC-equipped vehicles. One of the first such studies was carried out by Stern et al. 2018~\cite{stern2018dissipation}, where 21 vehicles drove on a circular track, following a setup similar to~\cite{sugiyama2008traffic,tadaki2013phase}. Initially, no ACC systems are activated and instability is clearly observed. However, when one of the vehicles engages its ACC controller with a high time gap, traffic flow became noticeably smoother. The study demonstrates that even with just 5\% of vehicles using ACC, stop-and-go waves could be significantly dissipated.

Contradicting these findings, other experiments have questioned the effectiveness of ACC systems in improving string stability. Makridis et al., 2018~\cite{makridis2018estimating} measure the reaction times of ACC vehicles and find them comparable to those of human drivers. Based on this, they predict a negative impact of ACC controllers on string stability and traffic flow. A few years later, the same researchers tested this hypothesis by conducting experiments at AstaZero in 2020 in Sweden~\cite{makridis2020empirical}. Their study reveals that the reaction times of the ACC vehicle ranged from 1.7 to 2.5 seconds, significantly longer than previously reported values. Testing five ACC-equipped vehicles from four different manufacturers, they observe clear signs of string instability. These contradictory results, along with the increasing prevalence of ACC vehicles, have sparked further research in the field, leading to a series of scientific investigations.

In 2019, Gunter et al.~\cite{gunter2019model} tested a 2015 luxury electric vehicle equipped with a commercial ACC system. By collecting data on velocity, relative velocity, and vehicle spacing, they calibrate an \emph{Optimal Velocity Relative Velocity} (OVRV) car-following model for both minimum and maximum following distance settings. Their analysis highlights that the best-fit models exhibited string instability. In a subsequent 2020 study, Gunter et al. seek to validate this finding empirically without relying on a car-following model. They test seven different ACC vehicles, collecting data from over 1,200 miles of car-following with ACC engaged \cite{gunter2020commercially}. Their results indicate that all tested vehicles, under all tested settings, are string unstable. In an eight-vehicle platoon, a 6 mph disturbance grew to 19 mph, ultimately causing the last vehicle in the platoon to drop below the minimum operational speed of its ACC system.

Following this, in 2021, Ciuffo et al.~\cite{ciuffo2021requiem} tested ten ACC-equipped vehicles at the ZalaZONE proving ground in Hungary. Their findings confirm the previously observed string instability of commercial ACC systems. The authors propose functional requirements to ensure the string stability of ACC-equipped vehicles, warning that without these improvements, ACC systems could negatively impact traffic safety and energy consumption.

Further supporting these conclusions, Makridis et al.~\cite{makridis2021openacc} introduce the \emph{OpenACC} database, which compiles data from several car-following experiments to analyze the properties of commercial ACC systems. Their analysis highlights that string stability is strongly influenced by the time-gap setting of ACC systems. While short time-gap settings led to unstable behaviour, longer time gaps promoted string stability.

\subsection{Stability of ACC Systems in Field Experiments}
\label{empircal_observatiosn}
Testing ACC systems in real-world field experiments is a challenge because of difficulties in tracking data, safety concerns, and maintaining platoons due to vehicles cutting in. Nevertheless, some studies with ACC systems in real-world scenarios have been conducted so far.

Fancher et al. 1998~\cite{fancher1998intelligent} test ACC driving behaviour with 108 participants. Viti et al. 2008~\cite{viti2008driving} analyse 20 Volkswagen Passats for six-month and found that the speeds and headway variance are lower with ACC active, but that the average headway increases. Schakel et al. 2017~\cite{schakel2017driving} conduct a study with eight participants using four different ACC car models and found that ACC controllers increases spacing and time headway in saturated conditions compared to human driving, suggesting a potential capacity decrease.

These mentioned studied test ACC systems empirically, but without focusing on string stability. 
One of the first real-world experiments that addresses string stability was conducted by Knoop et al.\ in 2019~\cite{knoop2019platoon}. In their study, seven vehicles equipped with ACC systems and lane-centering systems travel as a platoon on public roads for almost 500 km. Their investigation reveal that maintaining the platoon was challenging due to aggressive lane changes by surrounding human-driven vehicles. Additionally, platoons longer than three to four vehicles become too unstable, resulting in collision risks under oscillatory conditions.

Another empirical experiment was conducted by Li et al.\ in 2021~\cite{li2021car}, using a three-vehicle platoon with a human-driven lead vehicle followed by two ACC vehicles. They observe that the vehicles were string unstable, with oscillation amplitudes escalating rapidly in longer platoons. The study emphasise that this behaviour largely depended on factors such as time gap settings, speed levels, and the behaviour of the lead vehicle.

The experiments conducted by He et al.~\cite{he2020energy} are worth mentioning. In a real-world scenario, they track a platoon of five vehicles consisting of a human-driven lead vehicle, three ACC-equipped vehicles in the middle, and a human-driven vehicle at the end. Their findings show that ACC system have a negative impact on both stability and energy efficiency.

One of the largest field experiments was conducted as part of the CIRCLES project~\cite{hayat2022holistic}. In 2023, they evaluate the impact of 100 vehicles equipped with longitudinal control on traffic flow. In one study, a reinforcement learning-based (RL) ACC controller is introduced and tested within the fleet~\cite{jang2025reinforcement}. The study demonstrated that the ACC controller improve traffic flow and reduce the emergence of stop-and-go waves.

\subsection{Stability of Cooperative ACC Systems}
Cooperative Adaptive Cruise Control (CACC) systems enhances traditional ACC systems by integrating vehicle-to-vehicle (V2V) or vehicle-to-infrastructure (V2I) communication. This allows vehicles to share real-time information such as speed, acceleration, and intended maneuvers, enabling tighter coordination compared to ACC, which relies solely on onboard sensors. While the literature on ACC presents mixed findings—some studies report positive effects on collective stability, others negative—there is a broad consensus regarding CACC-equipped vehicles. Research consistently demonstrates that CACC systems outperforms ACC systems in terms of collective stability, with positive impacts on traffic flow and safety.

One of the first simulation-based studies on CACC stability was presented by van Arem et al., in 2006~\cite{van2006impact}. Their car-following model for CACC vehicles demonstrates improved traffic-flow stability, particularly in highway merging scenarios where the number of lanes decreases from four to three. The simulations highlighted CACC’s ability to mitigate disruptions, enhancing overall traffic efficiency. Building on this, Schakel et al., 2010~\cite{schakel2010effects} introduce a modified version of the \emph{Intelligent Driver} model (IDM+), tailored for CACC. Their results show that CACC enhances traffic stability by rapidly damping shockwaves—sudden variations in traffic speed or density—thus preventing their propagation through the traffic stream.

The advantages of CACC extend beyond simulations, with numerous experiments reinforcing its stability benefits. In 2004, de Bruin et al.~\cite{de2004design} conducted one of the first field tests with three communicating vehicles. They observe anticipatory braking actions enabled by V2V communication, which smoothed traffic flow and suggested improved stability. Bu et al., 2010~\cite{bu2010design} further advanced this work by testing two vehicles with wireless communication in the field and three on an experimental ground. Their results show a significant reduction in time gaps, from the typical ACC range of 1.1 to 2.2 seconds to just 0.6 seconds with CACC, indicating enhanced responsiveness and stability due to communication.
Public road testing by Milanes et al., 2013~\cite{milanes2013cooperative} with four CACC-equipped vehicles provides further evidence. They report improvements in response time and string stability compared to traditional ACC systems. This was followed by a comparative study by Milanes and Shladover, 2014~\cite{milanes2014modeling}, who test four vehicles equipped with both ACC and CACC systems. Their findings show that ACC systems exhibited string instability, whereas CACC systems are string stable.

Gong et al., 2019~\cite{gong2019cooperative} offer a nuanced perspective by comparing different CACC strategies against ACC. They evaluate scenarios with no communication (ACC), communication limited to the vehicle directly ahead, and communication with multiple vehicles in the platoon. Their results reveal that even local communication—connecting a vehicle only to its immediate predecessor—yielded substantial stability improvements over ACC. This suggests that full platoon-wide cooperation is not necessary to achieve significant benefits, enhancing CACC’s practicality for real-world deployment.
Xing et al., 2018~\cite{xing2018smith} propose a CACC algorithm to compensate for time delay and disturbances in transmission dynamics, such as wind gusts and friction. The system satisfies string stability with a time gap of $0.1$~seconds. Furthermore, they demonstrate the algorithm's effectiveness through simulations involving multiple vehicles and a platoon experiment using two real cars under controlled conditions. In 2019~\cite{xing2019compensation}, they extend this methodology to additionally compensate for inter-vehicle communication delays by leveraging a bidirectional V2V communication topology.
Zhang et al., 2023~\cite{zhang2023memory} formulate a memory-anticipation control strategy to address communication and time delays. The platooning system implementing this method demonstrates string stability, provided that the clocks in the transmitters and receivers of the communication units are synchronised across all participating vehicles.

While CACC technology shows promising results, the widespread adoption of CACC-equipped vehicles on roads would necessitate substantial infrastructure and regulatory modifications. Furthermore, some experts predict that the implementation of connected vehicles on the roads may take considerably longer than initially anticipated~\cite{raposo2019future}.






\section{Discussion}
In this contribution, we presented a comprehensive analysis of collective stability for ACC systems. We first introduced the concept of local, string, damped and overdamped stability theoretically, and then proceeded with the illustration of these concepts in simulations of car-following models, in experiments with ACC systems, and of real-world observations.
The overwhelming conclusion from these analysis is that current commercial ACC systems can exhibit instability phenomena due to manufacturers prioritizing smooth and comfortable driving experiences. Having a higher response time 
and no anticipation mechanism, they often perform worse than human drivers in maintaining stable traffic flow~\cite{makridis2018estimating}. To address this issue, one promising direction highlighted in this paper is the integration of connectivity. By enabling V2V communication, CACC systems facilitate the exchange of real-time information, allowing for reduced time gaps and faster reaction times. However, the widespread adoption of CACC remains limited in practice, underscoring the need for interim solutions, such as refining ACC controllers to prioritize stability without sacrificing safety or comfort. 
The aim is to develop robust autonomous systems that can compensate for different types of disturbances, whether delays, latency or random fluctuations related to environmental conditions or the accuracy of measurement sensors \cite{khound2023unified}.
Another promising approach in this regard is the implementation of reinforcement and deep reinforcement learning~\cite{jang2025reinforcement,das2021d}. 
Recent approaches aim notably at combining reinforcement strategies in CACC systems \cite{jiang2022reinforcement,peng2022collaborative}. 
These different techniques will undoubtedly improve the performance of adaptive cruise control systems in the coming years.





\section*{ACKNOWLEDGMENT}

The authors acknowledge the German Research Foundation (DFG, grant number 546728715).


\bibliographystyle{IEEEtran} 
\bibliography{biblio.bib}

\end{document}